\documentclass[sigconf]{acmart}
\AtBeginDocument{%
  }

\setcopyright{acmlicensed}
\copyrightyear{2026}
\acmYear{2026}
\acmDOI{XXXXXXX.XXXXXXX}
\acmConference[Conference acronym 'XX]{Make sure to enter the correct
  conference title from your rights confirmation email}{June 03--05,
  2026}{Woodstock, NY}
\acmISBN{978-1-4503-XXXX-X/2018/06}

\begin{document}

\title{LASER: An Efficient Target-Aware Segmented Attention Framework for End-to-End Long Sequence Modeling}

\author{Tianhe Lin*, Ziwei Xiong*, Baoyuan Ou, Yingjie Qin, Lai Xu\\ Xiaocheng Zhong, Yao Hu, Zhiyong Wang, Tao Zhou, Yubin Xu, Di Wu}

\affiliation{%
 \institution{Xiaohongshu Inc.}
  \city{Shanghai}
  \country{China\\
  \{xiongziwei,oubaoyuan,yingjieqin,xulai,mingcheng1,xiahou,sunzhenghuai,zhoutao3,xuyubin1\}@xiaohongshu.com, thlin20@fudan.edu.cn, wudi.xjtu@gmail.com
  }
}

\thanks{\textsuperscript{*}These authors contributed equally to this work.}

\renewcommand{\shortauthors}{Lin et al.}

\begin{abstract}
{Modeling ultra-long user behavior sequences is pivotal for capturing evolving and lifelong interests in modern recommendation systems. However, deploying such models in real-time industrial environments faces a strict ``Latency Wall'', constrained by two distinct bottlenecks: the high I/O latency of retrieving massive user histories and the quadratic computational complexity of standard attention mechanisms. To break these bottlenecks, we present \textbf{LASER} (\textbf{L}ong-sequence \textbf{A}ttention with \textbf{S}egmented \textbf{E}fficient \textbf{R}epresentation), a full-stack optimization framework developed and deployed at Xiaohongshu (RedNote). Our approach tackles the challenges through two complementary innovations: (1) System efficiency: We introduce SeqVault, a unified schema-aware serving infrastructure for long user histories. By implementing a hybrid DRAM-SSD indexing strategy, SeqVault reduces retrieval latency by 50\% and CPU usage by 75\%, ensuring millisecond-level access to full real-time and life-cycle user histories. (2) Algorithmic efficiency: We propose a Segmented Target Attention (STA) mechanism to address the computational overhead. Motivated by the inherent sparsity of user interests, STA employs a sigmoid-based gating strategy that acts as a silence mechanism to filter out noisy items. Subsequently, a lightweight Global Stacked Target Attention (GSTA) module refines these compressed segments to capture cross-segment dependencies without incurring high computational costs. This design performs effective sequence compression, reducing the complexity of long-sequence modeling while preserving critical signals. Extensive offline evaluations demonstrate that LASER consistently outperforms state-of-the-art baselines. In large-scale online A/B testing serving over 100 million daily active users, LASER achieved a 2.36\% lift in ADVV and a 2.08\% lift in revenue, demonstrating its scalability and significant commercial impact.}
\end{abstract}

\ccsdesc[500]{Information systems~Recommender systems}

\keywords{Long Sequence Modeling, Industrial Recommenders, Click-Through Rate Prediction}

\newcommand{\add}[1]{\textcolor{blue}{#1}}

\maketitle
\setlength{\textfloatsep}{1pt plus 2.0pt minus 2.0pt}
\setlength{\floatsep}{5pt plus 0.5pt minus 0.5pt}
\setlength{\abovecaptionskip}{2pt plus 0.5pt minus 0.5pt}
\setlength{\belowcaptionskip}{1pt plus 1pt minus 1pt}

\section{Introduction}
In modern recommendation systems, modeling user behavior sequences has become a cornerstone for capturing dynamic user interests and enhancing personalization~\cite{he2023survey, pan2026survey, hu2025practice}. 
As Xiaohongshu\footnote{An influential lifestyle platform with over 100 million users: www.xiaohongshu.com} (RedNote) evolves into a comprehensive content ecosystem encompassing multiple modalities and user interactions, user histories often consist of thousands of notes spanning diverse scenarios. 
Comprehensive modeling of these long sequences can significantly improve ranking performance and recommendation diversity~\cite{sun2019bert4rec,kang2018self}. 
However, due to computational constraints and limitations in data infrastructure, Xiaohongshu's existing systems have traditionally relied on either short sequences~\cite{zhou2018deep} or two-stage paradigms~\cite{chang2023twin,si2024twin,pi2020search}. 
The former performs end-to-end modeling via attention mechanisms~\cite{vaswani2017attention} over sequences of approximately $10^2$ in length, while the latter first retrieves the top-k notes (typically $k \approx 10^2$) most relevant to the current candidate from the original ultra-long sequence, followed by short-sequence modeling over this subset~\cite{xu2025gist}.
However, the retrieval stage often relies on hard filtering or heuristic similarity, leading to irreversible information loss and failing to capture lifelong interests.

In real-world recommendation scenarios, user decisions are not shaped by isolated recent behaviors, but rather by long-term, evolving patterns across different scenarios and interactions. Short sequences inherently truncate user footprints and lose critical signal diversity and temporal depth. 
Prior two-stage long-sequence pipelines extend the context length but introduce cascading information loss due to the separated retrieval and modeling.

Consequently, the limitations of previous sequence modeling approaches become particularly evident in advertisement ranking (click-through rate prediction) tasks, where accurate prediction depends on the model's ability to identify subtle correlations between candidate advertisements and the user's extensive historical footprints which encompass content interactions, search actions, product engagements, and multi-scenario behaviors accumulated throughout the user's life cycle.

Recently, complex attention-based architectures have demonstrated immense potential in sequence modeling~\cite{zhai2024actions,chai2025longer}, but industrial systems must operate under strict latency and computational constraints~\cite{xia2023transact, guan2025make}. This challenge is twofold. (1) System I/O: fetching thousands of behaviors creates massive disk IO, network, and serialization overhead. (2) Model compute: standard self-attention ($O(L^2)$) is computationally prohibitive for real-time serving.
This reality makes end-to-end modeling of ultra-long sequences a formidable challenge.
Although some research~\cite{chai2025longer,lyu2025dv365} focuses on enhancing model capabilities through architectural innovations, the design of sequence modeling must still strike a balance between performance and efficiency since extending historical sequences directly leads to a drastic increase in storage, bandwidth, and computational loads during distributed training.

In this work, we present LASER (\textbf{L}ong-sequence \textbf{A}ttention with \textbf{S}egmented \textbf{E}fficient \textbf{R}epresentation), a novel framework developed and deployed at Xiaohongshu to enable end-to-end ultra-long sequence modeling at scale.
LASER is designed to address three core challenges in practical large-scale deployment: \textbf{(1) Efficient access} to full life-cycle, multi-scenario user histories. \textbf{(2) Effective attention architecture} tailored for real-time, target-aware relevance modeling. \textbf{(3) Computational efficiency} under strict latency budgets.

To achieve this, we introduce a unified long sequence service named SeqVault that provides real-time  access to user behavior sequences with side information across all scenarios and stages of user engagement. 
Built upon this infrastructure, LASER employs a novel end-to-end attention mechanism over user sequences.
First, a fine-grained target-aware compression module performs low-dimensional, segmented target attention over long sequences, enabling noise-resilient compression via sigmoid-weighted aggregation within fixed-length segments.
Second, a global stacked target attention module operates over the compressed sequence to model cross-segment dependencies with minimal overhead. 
This ``compress-then-refine'' design reduces computational cost, while preserving critical semantic signals.

Deployed in production as part of Xiaohongshu's ranking system, LASER replaces previous short-sequence attention modules and integrates seamlessly with advanced feature interaction modules such as RankMixer~\cite{zhu2025rankmixer}.
Offline evaluations show consistent improvements in AUC and online A/B testing demonstrates significant gains in two key business metrics. 
Beyond immediate performance gains, LASER lays the foundational infrastructure for future multi-scenario unified modeling, marking a step forward in the pursuit of scalable, end-to-end sequence modeling in industrial recommendation systems. Overall, the contributions of this work are summarized as follows:
\begin{itemize}
    \item A production-validated system for real-time, end-to-end long sequence modeling at scale.
    \item A segmented target attention paradigm that preserves effectiveness and efficiency in long user sequence modeling.
    \item Demonstrated impact in an influential business scenario at Xiaohongshu, affecting over 100 million daily active users.
\end{itemize}

\section{Related Work}
\subsection{Short Sequence Modeling}
Modeling user behavior sequences in recommendation systems has long been recognized as a key component for capturing user interests~\cite{pan2026survey,he2023survey}. 
Early works typically operate on short-term user histories due to system constraints and latency budgets. 
Deep Interest Network (DIN)~\cite{zhou2018deep} pioneers the use of attention mechanisms in recommender systems, allowing the model to focus on items relevant to the candidate item.
To better capture sequential signals, subsequent studies have incorporated attention-augmented GRUs~\cite{zhou2019deep}, or employed self-attention mechanisms~\cite{chen2019behavior,kang2018self,sun2019bert4rec}.
Further refinements have explored richer inductive biases, including leveraging historical sessions~\cite{feng2019deep} and capturing diverse user interests through multi-interest modeling~\cite{li2019multi,10.1145/3477495.3532073}.

Despite their architectural sophistication, most of these models are designed under the assumption of relatively short input sequences (typically $\le100$ items), neglecting scalability and limiting their practical applicability to long sequences in real-time systems.

\subsection{Long Sequence Modeling}
\begin{figure*}[ht]
  \centering
  \includegraphics[width=0.86\linewidth]{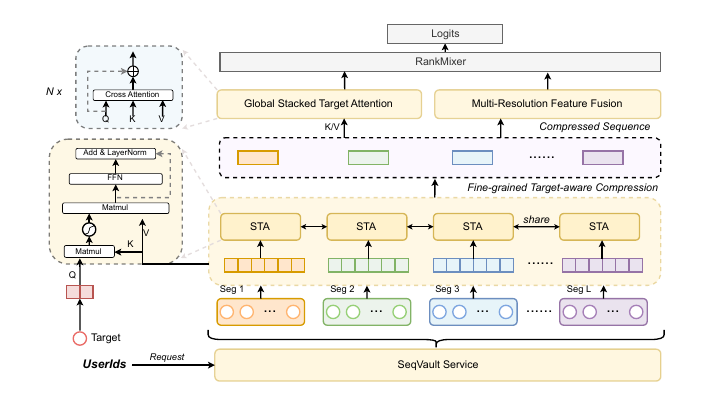}
  \caption{Overview of our LASER framework. The architecture comprises four key components: (1) The SeqVault service efficiently retrieves real-time user behavior sequences and side information. (2) The Segmented Target Attention (STA) module performs hierarchical processing, enabling sequence compression while preserving local patterns. (3) The Global Stacked Target Attention (GSTA) module conducts fine-grained interaction modeling on the compressed sequence. (4) The final representation consists of the attention output and the embeddings obtained by the multi-resolution feature fusion module, and is fed into RankMixer for CTR prediction.}
  \Description{A description of the laser figure.}
  \label{fig:overview}
\end{figure*}
Apart from scaling up dense parameters~\cite{zhang2024wukong,zhu2025rankmixer,yu2025hhfthierarchicalheterogeneousfeature}, extending user sequences represents a complementary and promising direction in recommendation systems~\cite{guan2025make}. 
Both research and industry have demonstrated growing interest in scaling up sequence length beyond the traditional short-term window.
One line of work focuses on two-stage retrieval frameworks. For example, SIM~\cite{pi2020search} first retrieves top-k relevant historical behaviors via hard search (by rule) or soft search (by embedding similarity) before applying attention to the selected sequences.

The following works solve the inconsistency between GSU and ESU by adopting an identical target-behavior relevance metric~\cite{chang2023twin,si2024twin}.
Another line of work chooses offline embedding approaches that trade off real-time freshness for the ability to capture ultra-long sequential patterns~\cite{yang2023empowering,lyu2025dv365}.
Typically, an offline foundational model is pretrained on ultra-long sequences, after which the condensed embeddings are leveraged by downstream recommendation systems.
Distinct from the aforementioned two categories, recent efforts have focused on modeling long sequences directly~\cite{zhai2024actions,chai2025longer,guan2025make}, yet this area remains largely under-explored.

\section{Method}

\subsection{Problem Formulation}
In industrial recommendation systems, the CTR prediction task in the ranking stage aims to accurately estimate the probability that a given user will click a specific item from a small candidate set generated by the recall phase.
We formulate the CTR prediction task as a conditional probability estimation problem. 
Given a user $ u $ with historical behavior sequence $ \mathcal{H} = \{h_1, h_2, ..., h_L\} $ and a target item $ t $, our objective is to learn a function $ f(\cdot) $ that maximizes the conditional probability:
\begin{equation}
P(\text{click}=1 | u, \mathcal{H}, t) = \sigma(f(u, \mathcal{H}, t; \theta))
\end{equation}
where $ \sigma(\cdot) $ denotes the sigmoid function, $ \theta $ represents model parameters, $ h_n \in \mathbb{R}^{d} $ is the $ d $-dimensional embedding of the $ n $-th historical interaction, and $ L $ denotes the sequence length.
The optimization goal is to minimize the binary cross-entropy loss:
\begin{equation}
\mathcal{L} = -\frac{1}{|\mathcal{D}|}\sum_{(u,\mathcal{H},t,y)\in\mathcal{D}} \left[ y\log \hat{y} + (1-y)\log(1-\hat{y}) \right]
\end{equation}
where $ \mathcal{D} $ denotes the training dataset, $ y \in \{0,1\} $ is the ground truth label, and $ \hat{y} = P(\text{click}=1 | u, \mathcal{H}, t) \in (0,1)$.

\subsection{Overview of LASER Framework}
Before diving into the detailed architecture, we first provide a brief overview of our LASER framework. 
As illustrated in Figure~\ref{fig:overview}, LASER integrates four key components. First, to enable real-time and customizable long-sequence modeling, we build a unified SeqVault service that exploits DRAM-SSD hybrid storage to offer highly efficient sequence access (detailed in Section \ref{subsec:seqvault}).
By leveraging centralized storage of user behavior sequences, our SeqVault service enables long-sequence training with 10x lower resource consumption while maintaining training and inference consistency.

Based on our SeqVault infrastructure, we introduce a novel and efficient model for end-to-end modeling of long user behavior sequences.
LASER employs a ``compress-then-refine'' strategy that strikes a balance between computational cost and performance.
The input sequence is partitioned into $ L' $ segments. For each segment $ S_i $, a low-dimensional Segmented Target Attention (STA) computes interaction scores between the target item and historical behaviors using a shared-parameterized query-key-value transformation, and aggregates each segment into one token $ \mathbf{s}_i $.
The compressed segments $ {\mathbf{s}_1, \mathbf{s}_2, ..., \mathbf{s}_{L'}} $ are then fused into a compact sequence $ \mathcal{H}' \in \mathbb{R}^{L' \times d} $. 

A lightweight Global Stacked Target Attention (GSTA) operates on $ \mathcal{H}' $ to model cross-segment dependencies, generating a refined representation $ z \in \mathbb{R}^{d} $. 
To simultaneously model long-term and short-term representations, the final representation $ z $, along with max-pooled features and recent $ r $ segments, are used as inputs to the subsequent feature interaction layer for CTR prediction.

\subsection{SeqVault Service}\label{subsec:seqvault}
 \begin{figure}[t]
     \centering
     \includegraphics[width=\linewidth]{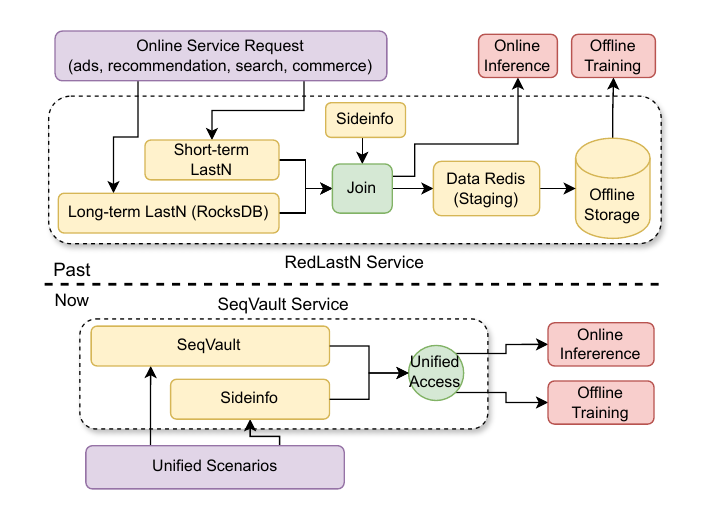}
     \caption{Comparison of traditional fragmented LastN infrastructures (past) with the unified SeqVault service (now), showcasing resource-efficient, real-time long sequence modeling and 10× cost reduction through centralized sideinfo management.}
     \Description{SeqVault}
     \label{fig:SeqVault}
 \end{figure}

Figure~\ref{fig:SeqVault} shows the evolution from the last-generation sequence service (RedLastN) to SeqVault. In the past, user behavior features from different scenarios had to go through an online procedure to be filled into the short-term lastN and long-term lastN as real-time and offline sequences accordingly. The short-term lastN is backed by an internal data service platform for real-time service, and the long-term lastN is backed by RocksDB for persistence. Then these user actions and their side information are joined together for online inference. The joined output is also sent to staging storage and offline storage to serve offline training. This architecture introduces several problems: (1) the lengthy data pipeline slows down feature onlining, (2) high disk waste due to RocksDB's schema unawareness when storing user sequences involving only simple integral values, (3) high P99 tail latency when accessing the long-term lastN. 

With SeqVault, the above problems are addressed by 1) unifying features from different scenarios to remove the online ingestion, 2) storing all features in a schema-aware manner, and 3) exploiting DRAM-SSD hybrid indexing to minimize access latency.

More specifically, SeqVault adopts a two-level indexing strategy to improve performance and storage efficiency. First, SeqVault keeps a hash table in the memory and sequence storage on the disk, exploiting DRAM's low latency for indexing and disk's large capacity for storage. Second, schema-awareness of SeqVault makes it possible to effectively pack the input sequences in a space-efficient way instead of storing every type of data as a string. Finally, SeqVault does not have a compaction procedure that involves considerably large data read/write, but only runs a sequence-level merge job to actively merge on-disk sequences and in-memory updates. Together, these designs yield excellent performance and satisfactory reduction in disk usage. Crucially, SeqVault serves as the prerequisite enabler for LASER, reducing the data access latency budget to allow more room for complex model inference.
\subsection{Model Structure}

\subsubsection{Input}
The first step is to prepare the input tokens for the user behavior sequence and the target item, retrieved efficiently via our SeqVault service.
The side information for each item in the user history includes ID features, multi-modal embedding features, and similarity scores~\cite{xu2025gist}. 
We also include a recency feature, representing the time difference between the historical event and the request time, encoded as positional embeddings. 
Formally, we obtain the user sequence tokens $\mathcal{H} = \{h_1, h_2, ..., h_L\}\in \mathbb{R}^{L \times d}$ and the target item token $t \in \mathbb{R}^{d}$.

\subsubsection{Fine-grained Target-aware Compression}
Directly modeling ultra-long sequences by self-attention incurs quadratic computational costs and introduces significant noise. 
To address this, we introduce the \textbf{Segmented Target Attention (STA)} module, which performs dimensional reduction while preserving task-relevant information.

First, the input sequence is uniformly partitioned into $L'$ non-overlapping segments, and each segment contains $w$ consecutive tokens:
\begin{equation}
    [S_1,...,S_{L'}] = \text{split\_seq}(\mathcal{H}, w)
\end{equation}
This segmentation strategy serves a dual purpose: (1) \textbf{Complexity Reduction:} It decouples local pattern extraction from global dependency modeling, reducing the effective sequence length for subsequent layers; (2) \textbf{Noise Isolation:} It confines the influence of irrelevant behaviors within local windows.

For each segment $S_i$, we compute a target-dependent attention score. 
Specifically, let $t \in \mathbb{R}^{d}$ denote the embedding vector of the target item. 
The query vector is derived from the projection of $t$ through a learnable matrix $\mathbf{W}_q \in \mathbb{R}^{d \times d_q}$, while keys and values are obtained from linear transformations of the segment's tokens with $\mathbf{W}_k \in \mathbb{R}^{d \times d_q}, \mathbf{W}_v \in \mathbb{R}^{d \times d}$. The attention operation follows:
\begin{equation}
    Q = t\mathbf{W}_q, \quad K=S_i\mathbf{W}_k, \quad V=S_i\mathbf{W}_v
\end{equation}
\begin{equation}
    \mathbf{s}_i = \text{STA}(t, S_i) = \text{Sigmoid}\left(\frac{QK^\top}{\gamma}\right) V
\end{equation}
where the projection matrices are shared across segments to maintain consistency, and $\gamma$ is a scaling factor used to stabilize the attention scores for the sigmoid activation. 

\paragraph{Optimization Strategies.} We employ two critical designs to ensure efficiency and effectiveness:

\begin{itemize}
    \item \textbf{Low-dimension Projection:} We map queries and keys to a lower dimension $d_q \ll d$. This reduces the computational cost of the attention mechanism from $O(L \cdot d^2)$ to $O(L \cdot d \cdot d_q)$.
    
    \item \textbf{Sigmoid Gating as a Silence Mechanism:} A key innovation in LASER is replacing the standard softmax with sigmoid aggregation. 
    In standard softmax, the sum of the attention scores equals $1$, which forces the model to allocate probability mass even to completely irrelevant items, leading to \textit{noise amplification}. 
    In contrast, the sigmoid activation acts as an independent gate where the sum is not constrained. When a segment contains no relevant information, the sigmoid output approaches zero. This provides a silence mechanism, allowing the model to effectively prune noisy items from the computational graph.
\end{itemize}

The compressed segment representations $\{\mathbf{s}_i\}$ then undergo a feed-forward network (FFN) to enhance expressiveness:
\begin{equation}
    \mathbf{s}_i = \text{LN}\left(\text{FFN}\left(\mathbf{s}_i\right)+\mathbf{s}_i\right)
\end{equation}
Finally, the segments are fused into a compact sequence $\mathcal{H}' =[\mathbf{s}_1, ..., \mathbf{s}_{L'}]\in \mathbb{R}^{L' \times d}$, encapsulating task-relevant patterns at a segment-level granularity.

\subsubsection{Global Target Attention over Compressed Sequence}
After compression, the sequence length is reduced by a factor of $w$, allowing us to apply deep interaction layers at a relatively low cost.
We employ a \textbf{Global Stacked Target Attention (GSTA)} module to capture cross-segment dependencies.
Let $l$ denote the $l$-th layer ($l = 1,\dots, L_{stack}$), and $t_l$ denote the target vector updated by the $l$-th layer ($t_0 = t$).
The attention process is defined as:
\begin{equation}\label{eq:single_target_attn}
    \mathbf{a}_l = \text{softmax}\left(\frac{\left(t_{l-1} \mathbf{W}_q^l \right )\left(\mathcal{H}' \mathbf{W}_k^l\right)^\top}{\sqrt{d}}\right), \quad {o}_l = \mathbf{a}_l (\mathcal{H}' \mathbf{W}_v^l)
\end{equation}
The target vector is updated via residual connection: $t_{l} = t_{l-1} + o_l$. The final output $z = t_{L_{stack}}\in \mathbb{R}^{d}$ adaptively combines higher-order cross-segment signals.

\subsubsection{Multi-resolution Feature Fusion}
The final representation integrates three complementary views to handle the multi-faceted nature of user interests:
\begin{itemize}
    \item \textbf{Global Context ($\mathbf{z}$):} Captures long-term, cross-segment dependencies derived from the global attention.
    \item \textbf{Salient Features ($\text{max\_pool}(\mathcal{H}')$):} Preserves the strongest signals across the entire history.
    \item \textbf{Local Recency ($\mathbf{s}_{1},...,\mathbf{s}_{r}$):} Explicitly models the most recent $r$ segments to capture immediate user intent, which improves the accuracy of real-time user modeling.
\end{itemize}
These representations are concatenated with other features and fed into the feature interaction layers (RankMixer) for CTR prediction.

\subsection{Deployment-Oriented Optimization}
\subsubsection{Global Attention with Segmented Aggregation}
While the segmented target attention mechanism effectively captures local granular patterns, the naive implementation that iterates through segments $[S_1, \dots, S_{L'}]$ introduces computational redundancy. 
Specifically, since the transformation matrices remain invariant across the entire sequence, performing separate and small-scale matrix multiplications for each segment is inefficient. 
Instead of slicing the input embedding sequence $\mathcal{H} \in \mathbb{R}^{L \times d}$ immediately, we perform global linear transformations and attention scoring, followed by a reshaping operation to achieve segmentation. This approach leverages the parallel processing capabilities of GPUs and eliminates redundant calculations.

First, we compute the query vector $Q$, global keys $\mathcal{K}$, and global values $\mathcal{V}$ for the entire history sequence $\mathcal{H}$ in a single pass. Since $t$ is fixed, $Q$ is computed once:
\begin{equation}
    Q = t\mathbf{W}_q \in \mathbb{R}^{1 \times d_q}
\end{equation}
Simultaneously, we project the full sequence $\mathcal{H}$ to obtain the global key and value matrices:
\begin{equation}
    \mathcal{K} = \mathcal{H}\mathbf{W}_k \in \mathbb{R}^{L \times d_q}, \quad \mathcal{V} = \mathcal{H}\mathbf{W}_v \in \mathbb{R}^{L \times d}
\end{equation}
We then compute the global attention scores $\boldsymbol{\alpha}$ measuring the relevance of the target item to all historical tokens simultaneously:
\begin{equation}
    \boldsymbol{\alpha} = \text{Sigmoid}\left( \frac{Q \mathcal{K}^T}{\gamma} \right) \in \mathbb{R}^{1 \times L}
\end{equation}
This vectorized operation significantly reduces overhead compared to $L'$ separate operations.

To recover the segment-level representations, we reshape the global attention scores and value matrices based on the window size $w$ and segment count $L'$. We define the reshaping operator $\mathcal{R}(\cdot)$ which maps a tensor of length $L$ to dimensions $(L' \times w)$.
The global scores $\boldsymbol{\alpha}$ and values $\mathcal{V}$ are restructured as:
\begin{equation}
    \hat{\mathbf{A}} = \mathcal{R}(\boldsymbol{\alpha}) \in \mathbb{R}^{L' \times 1 \times w}, \quad \hat{\mathbf{V}} = \mathcal{R}(\mathcal{V}) \in \mathbb{R}^{L' \times w \times d}
\end{equation}
Finally, the compressed representations for all segments are obtained via a batched matrix multiplication, effectively performing the weighted sum within each window:
\begin{equation}
    \mathcal{H}' = \text{Squeeze}\left(\hat{\mathbf{A}} \times \hat{\mathbf{V}}\right) \in \mathbb{R}^{L' \times d}
\end{equation}
where $\times$ denotes batch matrix multiplication. This optimized formulation yields mathematically equivalent results to the naive iterative form but improves training throughput by maximizing parallelization and minimizing redundant operations.

\subsubsection{Communication Optimization}
In addition, due to the long sequence length, communication between cluster machines has become another bottleneck.
For training and online inference, we implement heterogeneous transmission optimization using Zstandard (ZSTD) compression, achieving a 40\% reduction in network bandwidth consumption by leveraging dictionary-based encoding for embedding vectors transmitted between CPU and GPU workers.

\subsection{Computational Complexity}
In this section, we provide a detailed breakdown of the computational cost in terms of floating-point operations (FLOPs). We use the same notations following previous sections: the sequence length $L$, embedding dimension $d$ and $d_q$, segment window $w$, and compressed length $L' = L/w$.
For the feed-forward network, $r$ denotes the expansion ratio ($d_{\text{ffn}} = r \cdot d$).

\subsubsection{LASER Complexity Analysis}
The computation is decomposed into two stages:

\paragraph{1. Fine-grained Compression (STA)} 
This stage operates on the full sequence. By utilizing low-dimension projection where $d_q \ll d$, we minimize the cost:
\begin{itemize}
    \item \textit{Projections ($\mathcal{H} \to K, V$):} The value projection costs $L d^2$, and the low-dimension key projection costs $L d d_q$. The query projection is negligible ($1 \cdot d \cdot d_q$).
    \begin{equation}
        C_{\text{proj}} \approx L d^2 + L d d_q
    \end{equation}
    \item \textit{Attention \& Aggregation:} Computing scores via dot product ($Q K^\top$) and aggregation ($\boldsymbol{\alpha} V$).
    \begin{equation}
        C_{\text{attn}} = L d_q + L d
    \end{equation}
    \item \textit{Feed-Forward Network (FFN):} Crucially, the FFN is applied \textbf{only} to the compressed sequence.
    \begin{equation}
        C_{\text{ffn}} = L' \cdot (2 r d^2) = \frac{L}{w} 2rd^2
    \end{equation}
\end{itemize}
Summing these, the total cost for STA is:
\begin{equation}
    C_{\text{STA}} \approx Ld(d + d_q) + \frac{L}{w}(2rd^2)
\end{equation}

\paragraph{2. Global Refinement (GSTA)} 
This stage operates exclusively on the compressed $L'$ segments. For $M$ stacked layers, including projections and attention:
\begin{equation}
    C_{\text{GSTA}} \approx M \cdot L' \cdot (2d^2 + 2d) \approx \frac{L}{w} 2M d^2
\end{equation}

\paragraph{Total LASER Cost.} 
Combining all components, we have
\begin{equation}
    C_{\text{LASER}} \approx L d^2 \left( 1 + \frac{d_q}{d} + \frac{2r + 2M}{w} \right) 
\end{equation}
Since $w$ is typically set to $10$, the most expensive terms (FFN and deep layers) are scaled down significantly. The overall complexity remains linear: $\mathcal{O}(L \cdot d^2)$.

\subsubsection{Comparative Analysis}
We compare LASER with self-attention (Transformer) and standard target attention.

\begin{itemize}
    \item \textbf{Self-Attention (Transformer):} Dominated by the pairwise attention calculation ($L \times L$) and four projections.
    \begin{equation}
        C_{\text{SA}} \approx 4L d^2 + 2L^2 d \quad (\text{Quadratic w.r.t } L)
    \end{equation}
    
    \item \textbf{Standard Target Attention:} A single shallow layer on the full sequence.
    \begin{equation}
        C_{\text{TA}} \approx 2 L d^2 + 2L d \quad (\text{Linear})
    \end{equation}
\end{itemize}

\paragraph{Quantitative Comparison.}
Consider a typical setting: $L=1000$, $d=128$, $d_q=32$, $w=10$, $r=4$, $M=2$.
\begin{itemize}
    \item \textbf{Self-Attention:} $\approx 3.2 \times 10^8$ FLOPs (dominated by $L^2 d$).
    \item \textbf{LASER:} $\approx 0.4 \times 10^8$ FLOPs.
\end{itemize}
LASER reduces the computational cost by nearly an order of magnitude compared to self-attention. 
More importantly, LASER's cost is comparable to a single-layer target attention ($\approx 0.3 \times 10^8$), yet it achieves deep semantic modeling capabilities through its hierarchical design. This confirms that LASER achieves high model performance with low computational cost.

\section{Experiments}
\subsection{Experimental Settings}
\subsubsection{Dataset} We evaluate our model on the CTR prediction task using a large-scale production dataset collected from Xiaohongshu's advertising system. The dataset comprises hundreds of millions of daily impression logs over a one-month period. Each sample corresponds to an ad exposure event and contains the following components: (1) user profile features, (2) candidate ad features, (3) contextual features, and (4) the user's historical behavior sequence from our SeqVault service. The label is binary, indicating whether the user clicked on the ad.

To ensure temporal validity and prevent data leakage, we adopt a strict time-based split: the training set contains 30 consecutive days of data, while the test set consists of the subsequent 3 days. User behavior sequences are constructed using events that occurred strictly before the timestamp of each prediction instance. The sequences incorporate user interactions from three major scenarios in Xiaohongshu: the dual-column feed stream, the single-column video stream, and the search scenario, providing a comprehensive view of user engagement.

\subsubsection{Implementation Details} 
For distributed training, we adopt the parameter server (PS) architecture, in which the parameter servers aggregate gradients and update global model parameters, while the workers perform parallel forward and backward computations. 
This training process is deployed on a cluster equipped with 32 NVIDIA L20 GPUs to ensure high computational efficiency.
During the entire training process, the per-GPU batch size is fixed at 2048.
And we use different optimizers for sparse and dense embeddings, where the sparse part is optimized with Adagrad, and the dense part is optimized with Adam ($\beta_1=0.9$, $\beta_2=0.9999$).

The maximum sequence length $L$ is set to 1000. For the segmented target attention module, we use a segment length $w=10$, resulting in $L'=100$ compressed segments. We set the number of attention heads to 2, and the query/key projection dimension $d_q$ in the compression module is set to 16 per head for computational efficiency, while the value dimension remains equal to the embedding dimension $d=128$.

\subsubsection{Baselines}
Our initial baseline is established using 100-length RocksDB sequences (RedLastN) from the online system, integrated with a modeling structure that combines target attention and self-attention mechanisms. 
To comprehensively evaluate the performance of our proposed model on the 1000-length SeqVault sequences, we select a series of state-of-the-art (SOTA) methods for comparative analysis. 
Specifically, the comparative models include DIN (target attention), HSTU, and Transformer (self-attention).
For a fair comparison, all baseline models are enhanced with a two-stage retrieval framework (GSU-ESU) operating on the user's full lifelong history (up to 10K items).

\subsubsection{Metrics}
Following standard practice in CTR prediction, we use the Area Under the ROC Curve (AUC) as the primary offline evaluation metric. AUC measures the model's overall ranking ability independent of threshold selection.
For online A/B testing, we report two key business metrics for the advertising system: revenue and the advertiser value (ADVV).

\subsection{Gains of SeqVault}
RedLastN (based on RocksDB) and SeqVault are both deployed in the cloud and replicated across regional zones. Table~\ref{tab:seqvault_resource} summarizes the reduction in resources of SeqVault (CPU cores per zone).

\begin{table}[ht]
  \caption{Resource reduction by SeqVault.}
  \label{tab:seqvault_resource}
  \begin{tabular}{ccccc}
    \toprule
    & \textbf{CPU cores} & \textbf{CPU usage} & \textbf{P99} & \textbf{Disk} \\
    \midrule
    RocksDB & 4327.13 & 16$\sim$24\% & 3$\sim$5ms & 140 TiB\\
    SeqVault & 3686.4 & $\sim$5\% & $\sim$ 2ms & \ 76.8 TiB\\
    Reduction & 14.8\% & 75\% &  $>$ 50\% & 45\% \\
    \bottomrule
  \end{tabular}
\end{table}

Due to the DRAM-SSD hybrid indexing structure, SeqVault reduces P99 latency by more than 50\% and also saves stalled CPUs during disk I/O. The schema-based storage enables compact data layout, significantly reduces disk usage, and makes sequence access faster, effectively reducing CPU usage and CPU cores.

\subsection{Offline Evaluation}
We conduct comprehensive offline experiments on a large-scale dataset from the Xiaohongshu Ads platform to evaluate the effectiveness of LASER against state-of-the-art sequence modeling baselines. 
All models are trained on the same dataset with identical features and evaluated using AUC as the primary metric.

As shown in Table~\ref{tab:perf}, we first establish a strong baseline using the production model with 100-length sequences. Extending the sequence length to 1,000 items and applying a standard target attention mechanism (DIN) yields a modest $+0.12\%$ AUC gain. This indicates the benefits of extending sequence length for CTR modeling, as longer sequences provide more contextual information.

More sophisticated architectures show clear advantages. HSTU and Transformer outperform standard target-attention methods, achieving gains of $+0.20\%$ and $+0.22\%$ respectively. This highlights that complex attention mechanisms have a distinct advantage in capturing complex dependencies within user behavior. 

Our proposed LASER framework achieves the highest performance with an AUC of 0.7826, corresponding to a $+0.24\%$ gain over the base model and outperforming all other baselines. The superior performance of LASER can be attributed to its synergistic design: the \textit{segmented target attention} efficiently distills relevant signals from the long sequence while mitigating noise, and the subsequent \textit{global stacked target attention} refines these compressed representations by modeling higher-order cross-segment dependencies. This ``compress-then-refine'' strategy effectively balances model capacity with computational feasibility for ultra-long sequences.

Overall, these results validate LASER's effectiveness and efficiency for modeling user behavior sequences, making it a compelling solution for industrial-scale long-sequence modeling.

\begin{table}[ht]
  \caption{Offline results on Xiaohongshu Ads platform.}
  \label{tab:perf}
  \begin{tabular}{cccc}
    \toprule
    \textbf{Method} & \textbf{AUC} & \textbf{AUC Gain} & \textbf{FLOPs} \\ 
    \midrule
    Base & 0.7802 & - & 1.3 $\times 10^7$ \\
    DIN & 0.7814 & +0.12\% & 3.3$\times 10^7$\\
    TWIN & 0.7810 & +0.08\% & - \\
    HSTU & 0.7822 & +0.20\% & 3.7$\times 10^8$\\
    Transformer & 0.7824 & +0.22\% & 3.6$\times 10^8$\\
    LASER & 0.7826 & +0.24\% & 4.0$\times 10^7$\\
    \bottomrule
  \end{tabular}
\end{table}

\subsection{Ablation Study}
To validate the contribution of each design choice in LASER, we conduct a systematic ablation study. We first examine the impact of removing specific architectural components, and then investigate the sensitivity of the model to the segment size hyperparameter.

\subsubsection{Impact of Model Components}
\begin{table}[ht]
  \caption{Ablation on components of LASER.}
  \label{tab:ablation}
  \begin{tabular}{cc}
    \toprule
    \textbf{Component} & $\Delta$\textbf{AUC(\%)}\\
    \midrule
    Sigmoid Aggregation & -0.03  \\
    FFN+LayerNorm & -0.08 \\
    LayerNorm & -0.04 \\
    Multi-resolution Fusion & -0.08 \\
    Recency embedding & -0.09 \\
    \bottomrule
  \end{tabular}
\end{table}

We evaluate the importance of key modules by removing or replacing them. The results are summarized in Table \ref{tab:ablation}.

We first examine the impact of the aggregation strategy in the compression stage. Replacing our proposed \textit{sigmoid aggregation} with the conventional softmax results in a $0.03\%$ performance drop. This indicates that sigmoid's independent scoring mechanism allows for more flexible, noise-resilient selection of relevant behaviors within each segment, avoiding the forced normalization of softmax, which can amplify irrelevant signals.
The -0.03\% drop with softmax validates our ``silence mechanism'' hypothesis: Softmax fails to filter out noise in irrelevant segments, whereas sigmoid successfully acts as a gate.

Next, we assess the components within the sequence compression block. Removing both the feed-forward network (FFN) and LayerNorm after segment aggregation causes significant degradation ($-0.08\%$). This underscores the importance of the FFN's non-linear transformation capability and LayerNorm's stabilization effect in learning effective segment-level representations. Removing LayerNorm alone results in a $0.04\%$ drop, highlighting its role in training stability.

We then evaluate our multi-resolution feature fusion strategy. When we remove the fusion of the \textit{max-pooled features} (which capture the most salient signals across the entire compressed sequence) and the explicit embeddings of the \textit{most recent $r$ segments} (which model short-term, fine-grained recency effects), performance decreases by $0.08\%$. This is the second-largest drop observed, demonstrating that complementing the global attention vector with these auxiliary views provides a more comprehensive behavioral representation, balancing long-term patterns with local temporal context.

Finally, the \textit{recency embedding}, which encodes the time difference between each historical behavior and the prediction request, proves to be the most critical single component. Its removal results in the largest performance drop ($-0.09\%$). This strongly suggests that temporal proximity is a paramount signal in user-interest modeling, as recent interactions typically carry greater predictive relevance for immediate decisions.

In summary, all proposed components contribute positively to LASER's performance. The ablation study confirms that our design choices are well-founded and collectively enable effective modeling of ultra-long user sequences.

\subsubsection{Impact of Segment Size}
The segment size $w$ in the segmented target attention module is a critical hyperparameter that controls the trade-off between information granularity and computational efficiency. We investigate the model's performance by varying $w \in \{5, 10, 20\}$ while fixing the total sequence length $L=1000$. The results are summarized in Table~\ref{tab:ablation_segment}.

\begin{table}[ht]
  \caption{Impact of segment size $w$ on performance and efficiency (with fixed $L=1000$).}
  \label{tab:ablation_segment}
  \begin{tabular}{cccc}
    \toprule
    \textbf{Segment Size} ($w$) & \textbf{AUC} & \textbf{Relative FLOPs} \\
    \midrule
    5 &  0.7823 & 1.5$\times$ \\
    \textbf{10 (Default)}  & \textbf{0.7826} & \textbf{1.0$\times$} \\
    20  & 0.7821 & 0.75$\times$ \\
    \bottomrule
  \end{tabular}
\end{table}

As observed in Table~\ref{tab:ablation_segment}, setting $w=10$ achieves the best performance. We analyze the behavior as follows:

\textbf{Small segment size ($w=5$).} Although a smaller window preserves finer granularity, it results in a longer compressed sequence ($L'=200$). This has two negative side effects. First, it significantly increases the computational load of the subsequent global stacked target attention module. Second, our model explicitly utilizes the last $r$ segments as recent features to capture short-term interests. With a small $w$, these $r$ segments cover a much narrower span of user history, potentially failing to capture sufficient recent context.
    
\textbf{Large segment size ($w=20$).} Increasing $w$ aggressively reduces the sequence length, improving efficiency. However, the performance drops significantly. An overly large window forces diverse user behaviors to be aggregated into a single vector, causing severe information loss. This prevents the global attention module from distinguishing specific user interests within the segment, thereby degrading the modeling of cross-segment dependencies.

Therefore, $w=10$ represents the optimal ``sweet spot'', providing sufficient granularity for the global attention to operate effectively while ensuring the recent segment features cover a meaningful temporal range without incurring excessive computational overhead.

\subsection{Scaling Analysis}
\begin{figure}[ht]
\centering
\includegraphics[width=\linewidth]{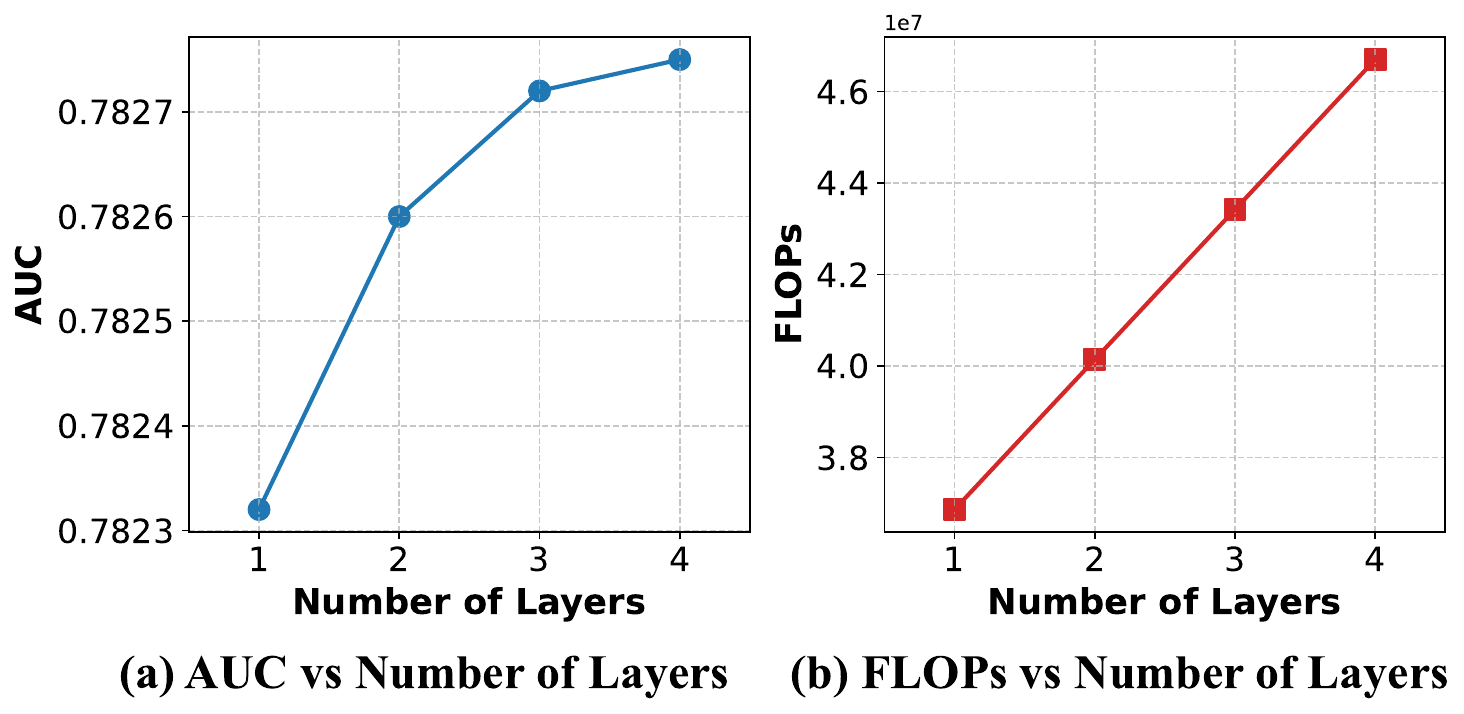}
\includegraphics[width=\linewidth]{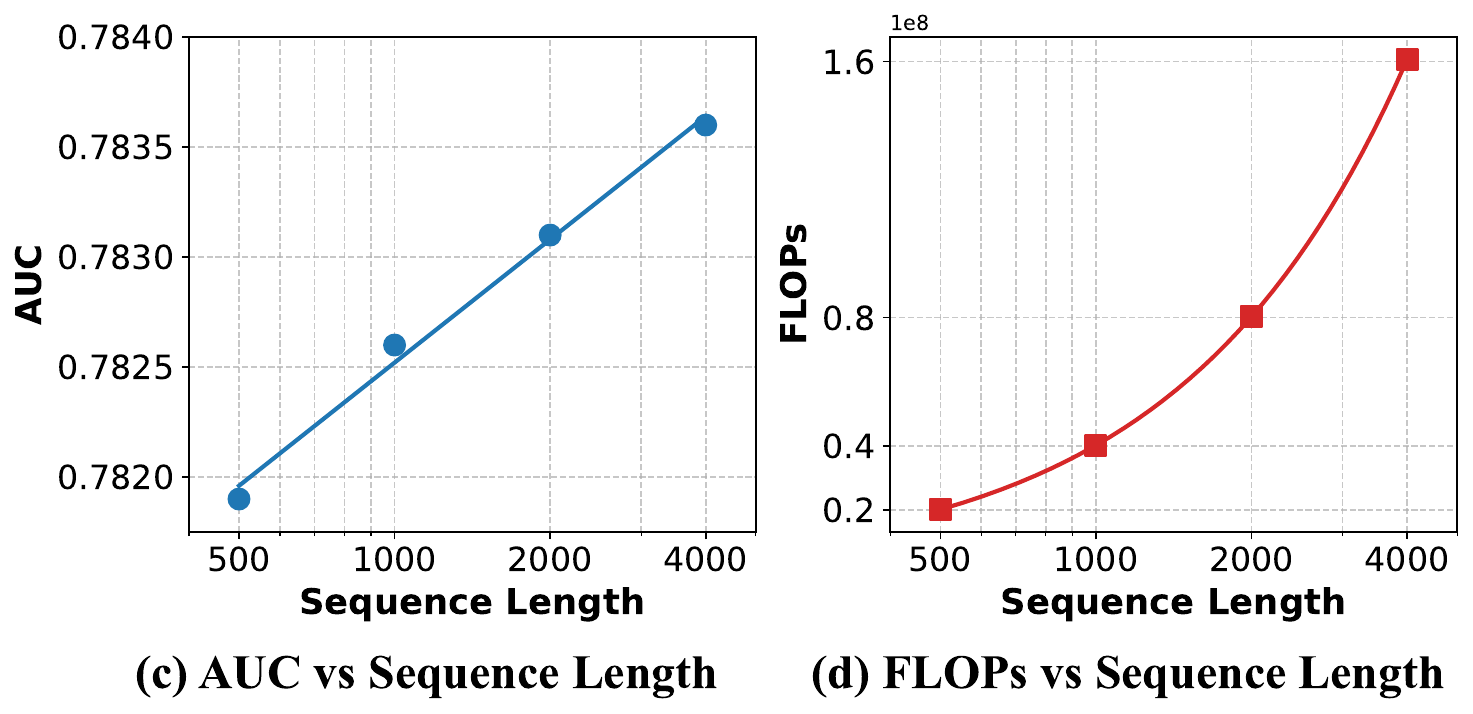}
\caption{Scaling analysis of LASER with varying number of layers and sequence lengths. (a) and (b) demonstrate that increasing model depth improves AUC with diminishing returns, while computational cost scales linearly. (c) and (d) show AUC follows a power-law relationship with increasing sequence length.}
\Description{Scaling analysis.}
\label{fig:scaling}
\end{figure}

To investigate the impact of model scale and sequence length on performance and practical feasibility, we conduct a systematic study focusing on two key dimensions: the number of GSTA layers and the input sequence length. The results are summarized in Figure~\ref{fig:scaling}.
\paragraph{Layer Scaling.}
We first examine the effect of model depth by varying the number of layers $M \in \{1, 2, 3, 4\}$. As shown in Figure~\ref{fig:scaling}(a), increasing the network depth leads to improved AUC; however, we observe significant diminishing returns. The transition from 1 to 2 layers yields the most substantial performance boost, whereas extending the model to 3 or 4 layers provides only marginal gains. Meanwhile, Figure~\ref{fig:scaling}(b) indicates that computational cost (FLOPs) grows linearly with the number of layers.
Based on this trade-off, we select two layers ($M=2$) as the optimal configuration. This choice draws a balance between performance benefits, inference efficiency and computational overhead. 
\paragraph{Length Scaling.}
We further test LASER with user sequences of length $L \in \{500, 1000, 2000, 4000\}$. Figure~\ref{fig:scaling}(c) reveals a clear power-law relationship where AUC improves as length increases, validating the importance of expanding sequence context. However, similar to the layer analysis, we observe diminishing returns beyond certain lengths. While longer sequences provide richer context, Figure~\ref{fig:scaling}(d) shows that FLOPs increase significantly due to the complexity of attention mechanisms.
From a deployment perspective, scaling to $L=1000$ yields a significant AUC gain over shorter sequences, while further extensions offer progressively smaller improvements relative to the increased resource consumption. Consequently, we adopt $L=1000$ alongside the 2-layer configuration to strike the best balance between performance gains and infrastructure costs.

\subsection{Online A/B Tests}
In this section, we present comprehensive online A/B testing results to validate the effectiveness of LASER on Xiaohongshu's advertising platform. We conducted a 7-day large-scale online experiment and  configured the experiment with strict traffic control to ensure statistical validity, allocating 10\% of total traffic to the experimental group.
As shown in Table~\ref{tab:lift}, the results demonstrate statistically significant improvements across key business metrics: \textbf{+2.36\%} increase in ADVV and \textbf{2.08\%} increase in revenue.

These gains represent meaningful business impact, as in our production environment, a sustained \textbf{0.5\%} improvement in either metric qualifies as a significant enhancement. Given the scale of Xiaohongshu's Ads platform, a 2.36\% lift in advertiser value translates to significant revenue growth, validating the commercial viability of our approach.

\begin{table}[ht]
  \caption{Online lift of LASER.}
  \label{tab:lift}
  \begin{tabular}{cccc}
    \toprule
    \textbf{ADVV} & \textbf{Revenue}\\
    \midrule
    +2.36\% & 2.08\% \\
    \bottomrule
  \end{tabular}
\end{table}

\section{Conclusion}
In this paper, we present LASER, a practical framework for end-to-end modeling of ultra-long user sequences, tackling the critical challenges of modeling ultra-long user behavior sequences at an industrial scale. By combining a unified SeqVault service with a novel hierarchical attention mechanism, LASER addresses the core challenges of sequence access efficiency and computational feasibility while maintaining strict latency constraints. Our SeqVault infrastructure enables real-time access to complete user behavior sequences spanning across multiple scenarios while reducing resource consumption by tenfold. At the model level, we introduce segmented target attention to achieve noise-resilient sequence compression, along with global stacked target attention for modeling cross-segment dependencies.

The system's effectiveness is validated through extensive offline experiments and large-scale online deployment at Xiaohongshu, where it achieves significant improvements in both prediction accuracy (AUC +0.24\%) and key business metrics (ADVV +2.36\% and Revenue +2.08\%). Beyond immediate performance gains, LASER establishes a foundational infrastructure for future advancements in long-sequence modeling, paving the way for even more comprehensive user understanding and personalized recommendations at scale.


\bibliographystyle{ACM-Reference-Format}
\bibliography{laser}


\end{document}